\documentclass[journal=jctcce,manuscript=article]{achemso}

\usepackage[T1]{fontenc}
\usepackage[version=3]{mhchem} % Formula subscripts using \ce{}
\usepackage{hyperref}
\usepackage{booktabs}

\author{Michael Brocidiacono}
\affiliation{Eshelman School of Pharmacy, University of North Carolina at Chapel Hill}
\email{mixarcid@unc.edu}\author{Brandon Novy}
\affiliation{Curriculum in Bioinformatics and Computational Biology, University of North Carolina at Chapel Hill}
\author{Rishabh Dey}
\affiliation{Eshelman School of Pharmacy, University of North Carolina at Chapel Hill}
\author{Konstantin I. Popov}
\affiliation{Eshelman School of Pharmacy, University of North Carolina at Chapel Hill}
\author{Alexander Tropsha}
\affiliation{Eshelman School of Pharmacy, University of North Carolina at Chapel Hill}

\title{Binding Free Energies without Alchemy}

\begin{document}

% \begin{tocentry}
% TODO: Add graphical table of contents entry
% \end{tocentry}

\begin{abstract}

Absolute Binding Free Energy (ABFE) methods are among the most accurate computational techniques for predicting protein-ligand binding affinities, but their utility is limited by the need for many simulations of alchemically modified intermediate states. We propose Direct Binding Free Energy (DBFE), an end-state ABFE method in implicit solvent that requires no alchemical intermediates. DBFE outperforms OBC2 double decoupling on a host-guest benchmark and performs comparably to OBC2 MM/GBSA on a protein-ligand benchmark. Since receptor and ligand simulations can be precomputed and amortized across compounds, DBFE requires only one complex simulation per ligand compared to the many lambda windows needed for double decoupling, making it a promising candidate for virtual screening workflows. We publicly release the code for this method at \url{https://github.com/molecularmodelinglab/dbfe}.

\end{abstract}

\section{Introduction}

A critical early step in drug discovery is finding small-molecule binders to proteins of interest. Structure-based virtual screening aims to speed up this process by computationally suggesting binders given the 3D structure of the target protein. In a virtual screen, a model scores millions or even billions of compounds for their ability to bind to the target; the top-scoring compounds are then suggested for experimental validation. In a typical screen, a ``docking'' algorithm first predicts a binding pose for each compound within the protein binding site, and then another algorithm is used to score the compound for its predicted binding ability.

While virtual screening is a promising alternative to expensive experimental high-throughput screens, it is currently far from reliable; current models only achieve modest enrichment of binders over nonbinders in a virtual screen \cite{brocidiacono_improved_2024, kolb_docking_2009, scior_recognizing_2012}. Traditional scoring methods such as AutoDock Vina \cite{trott_autodock_2010} and Glide \cite{friesner_glide_2004} use a physics-inspired energy function with a single frozen receptor conformation, neglecting the importance of the inherently dynamic nature of the system\cite{li_overview_2019}. While such methods are fast enough for use in large-scale screens, their accuracy is limited.

Recently, machine learning (ML) methods have attracted interest as scoring functions \cite{mcnutt_gnina_2021, francoeur_three-dimensional_2020, corso_diffdock_2022}. These utilize complex functions whose parameters are fit according to training data, and they can often achieve high accuracy on protein-ligand complexes that are similar to those seen in the training set. However, their success has been much more limited on novel protein targets that are dissimilar from those it has trained on \cite{yang_predicting_2020, volkov_frustration_2022, brocidiacono_bigbind_2023}.

On the other hand, absolute binding affinity (ABFE) calculations utilize rigorous statistical mechanics to predict the Gibbs free energy of binding. Several ABFE methods have been developed, most notably double decoupling (DD) \cite{gilson_statistical-thermodynamic_1997}. All such methods are computationally expensive due to the need to run many molecular dynamics simulations on ``alchemically modified'' states; however, they are highly correlated with experimental binding affinities. Thus, ABFE has emerged as the gold standard computational technique to predict protein-ligand binding affinities in the absence of experimental data about the protein target \cite{alibay_evaluating_2022, fu_meta-analysis_2022, aldeghi_accurate_2015}. While ABFE is far too slow to screen millions of compounds directly, it can be used to re-rank the top-scoring hits from an initial screen \cite{feng_absolute_2022, chen_enhancing_2023}. MD simulations for ABFE typically employ explicit waters, though ABFE with implicit solvent has been suggested as a faster alternative \cite{setiadi_tuning_2024}.

Between these two extremes lie several methods with intermediate accuracy and speed. They can be used to re-rank initial screening hits, possibly before a final ABFE run. Molecular Mechanics/Poisson Boltzmann (or Generalized Born) Surface Area (MM/PBSA or MM/GBSA) also uses molecular dynamics simulations to predict binding free energy, but does not simulate alchemical intermediates; it is thus faster, but at the expense of neglecting conformational entropy \cite{genheden_mmpbsa_2015}. Mining Minima avoids expensive molecular dynamics simulations altogether, and instead explicitly approximates partition functions as a sum over multiple harmonic energy wells \cite{chen_modeling_2010}. Finally, AlGDock approximates the binding free energy as an exponentially weighted sum of binding free energies to rigid receptors, which it can compute more easily by precomputing potential energy grids for each receptor conformation \cite{minh_alchemical_2020}.

In this work, we propose Direct Binding Free Energy (DBFE), a new implicit solvent ABFE method that utilizes only end-state simulation data; no alchemical intermediates are needed. DBFE requires only three simulations: the receptor-only, ligand-only, and receptor-ligand complex simulations. Since the receptor and ligand simulations can be precomputed and cached, the per-ligand cost reduces to a single complex simulation compared to the many complex lambda windows needed for DD. We evaluate DBFE on both host-guest and protein-ligand benchmarks, finding that it outperforms OBC2 DD on host-guest systems in correlation and performs comparably to OBC2 MM/GBSA on protein-ligand systems. Our results suggest that the primary accuracy bottleneck for implicit solvent methods on protein-ligand systems is the solvent model itself, rather than conformational entropy.

\section{Methods}

\subsection{Intuition}

Alchemical binding free energy methods seek to estimate the free energy of transforming a protein-ligand system from an initial potential energy function $U_0$ to a final energy function $U_1$. For absolute binding free energy calculations, $U_0$ is a decoupled state without any protein-ligand intermolecular interactions, and $U_1$ is the coupled state including all interaction terms. In principle, we could use methods such as the Bennett acceptance ratio (BAR) \cite{bennett_efficient_1976} to directly compute the binding free energy from samples from the two states (obtained via MD simulation). In practice, however, methods such as BAR require a certain amount of \textit{phase-space overlap} between the two states in order to return accurate estimates \cite{klimovich_guidelines_2015}; that is, at least some samples from $U_0$ should have a high probability of being sampled from $U_1$ (and vice-versa). Because phase-space overlap between the two end states is so low, alchemical methods resort to sampling from many intermediate energy functions $U_\lambda$, with $\lambda$ values spaced such that enough phase-space overlap exists between adjacent states. This requires many simulations, however, and is thus slow.

What is the source of the low phase-space overlap between the end states? In an implicit solvent, a state $x$ can be separated into the internal degrees of freedom of the ligand $x_L$, the internal degrees of freedom of the protein $x_P$, and the rigid-body transformation between them $\zeta$. There is often phase space overlap between $x_P$ in $U_0$ and $U_1$ (that is, proteins often sample holo-like states from apo simulations); similarly, there is phase space overlap between $x_L$ in the two states. We can easily introduce phase-space overlap in $\zeta$ via a restraining potential $U_\zeta$, a common practice in ABFE simulations \cite{clark_comparison_2023}. Thus, we argue that the low phase-space overlap between $U_0$ and $U_1$ is due primarily to the coupling between these degrees of freedom; the majority of samples from $U_0$ would have steric clashes between the protein and the ligand, causing a very high $U_1$ for that sample.

However, we can sample many possible permutations of $x_P$, $x_L$, and $\zeta$. Consider the decoupled energy function with a restraint on $\zeta$: $U_r(x) = U_P(x_P) + U_L(x_L) + U_\zeta(\zeta)$. We don't need to run molecular dynamics on this energy function to sample from its Boltzmann distribution; instead, we can run protein-only MD to sample $N_P$ samples from $U_P$ and ligand-only MD to get $N_L$ samples from $U_L$. We can then sample $N_\zeta$ rigid-body transformations, and then combine all these to produce $N_P N_L N_\zeta$ samples from $U_r$. This is enough samples that some will invariably not have steric clashes.

Attempting to use all these samples to get the free energy difference between $U_r$ and $U_1$ is still computationally intractable due to the need to evaluate both $U_r$ and $U_1$ on each sample. Spatial data structures such as KD-trees \cite{friedman_algorithm_1977}, however, allow us to quickly identify steric clashes in samples without evaluating the full energy function using fast nearest-neighbor searches. Using such a structure, we can quickly identify the subset of samples from $U_r$ without steric clashes. We can then use BAR to compute the free energy between $U_1$ and this subset. It is then easy to calculate the entropy difference between $U_r$ and the clash-free subset of $U_r$ using the number of samples that passed our filter. After a standard-state correction to account for the addition of $U_\zeta$, we can obtain the final free energy of binding.

\subsection{The Direct Binding Free Energy method}

We desire to compute the Gibbs free energy $\Delta G$ for the transformation $P + L \rightarrow PL$, where $P$ is a protein, $L$ is a small molecule ligand, and $PL$ is their complex. We can do this by computing the free energy difference between a system with a ``decoupled'' potential energy $U_0(x) = U_P(x_P) + U_L(x_L)$ and a system with the coupled potential energy $U_1(x) = U_{PL}(x)$. We assume we have run MD simulations of the protein, ligand, and protein-ligand complex; thus we have access to samples from $U_P$, $U_L$, and $U_{PL}$.

\begin{figure}[h]
\includegraphics[width=6in]{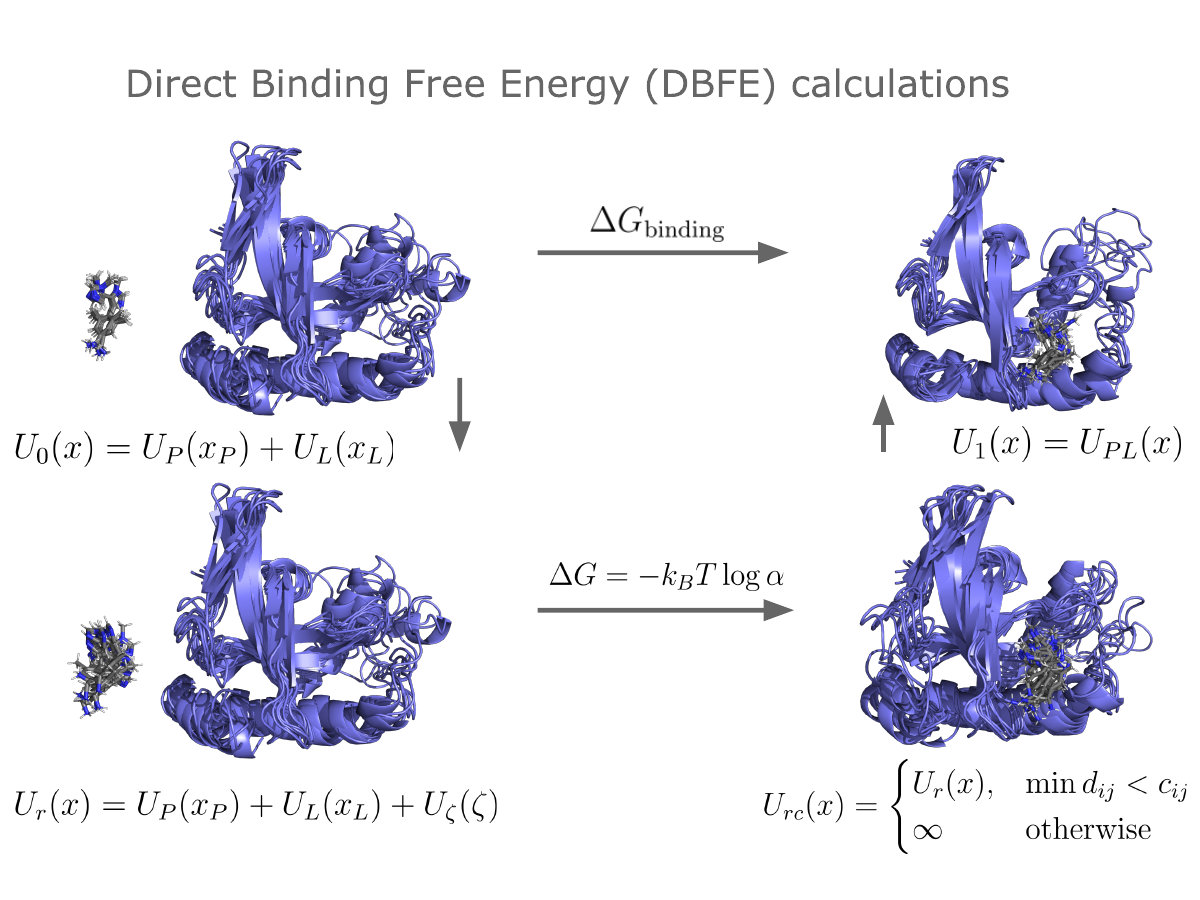}
  \caption{The full thermodynamic cycle used by DBFE.}
  \label{fig:therm_cycle}
\end{figure}

The thermodynamic cycle we use for this method is shown in Figure \ref{fig:therm_cycle}. In the first leg of the cycle, we seek to find the free energy of restraining the relative translation and rotation between the protein and the ligand. In this work, we define $\zeta$ to be the rigid-body transformation between the current ligand coordinates $x_L$ and the ligand coordinates of an arbitrary reference frame from the $PL$ simulation, after the pocket $\alpha$ carbons of the current frame have been aligned with this same reference frame. In this work, all ligand transformations account for molecular symmetry (i.e., equivalent atom permutations) via modified code from \texttt{spyrmsd} \cite{meli_spyrmsd_2020}. The restrained potential energy function is thus:

\begin{equation}
    U_{r}(x) = U_P(x_P) + U_L(x_L) + U_\zeta(\zeta)
\end{equation}

In this work, we define $U_\zeta$ as follows (though many other restraining potentials may also be used):

\begin{equation}
    U_\zeta(\zeta) = k_BT \left[ \frac{1}{2}(t - \mu)^T\Sigma^{-1}(t - \mu) - q^T MZM^T q \right]
\end{equation}

Here we have separated $\zeta = [ t, q ]^T$ into a translation vector $t$ and a rotation quaternion $q$. The first term is a multivariate normal distribution over translations with mean $\mu$ and covariance $\Sigma$. The second term is a Bingham distribution over rotations \cite{bingham_antipodally_1974}; the Bingham distribution is the analogue of the Gaussian on the unit quaternion sphere $S^3$ (and thus on the rotation group SO(3)), parameterized by an orthogonal matrix $M$ defining the principal axes and a diagonal matrix $Z$ controlling the concentration along each axis. This potential enables us to easily sample from $U_\zeta$ and analytically compute the free energy $\Delta G_{0\rightarrow r}$. We fit the parameters $M$, $Z$, $\Sigma$, and $\mu$ from the largest mode of the $PL$ trajectory (computed via the \texttt{MeanShift} function from \texttt{scikit-learn} \cite{fukunaga_estimation_1975, pedregosa_scikit-learn_2011}).

The second leg of the cycle corresponds to the KD-tree filtering step described above. We seek to find the free energy difference between $U_r$ and a \textit{conditional} energy function $U_{rc}$ that only has support on the samples without steric clashes.

\begin{equation}
    U_{rc}(x) =
\begin{cases}
U_r(x), & d_{ij} \geq c_{ij} \text{ for all atom pairs } (i,j)\\
\infty & \text{otherwise}
\end{cases}
\end{equation}

In order to compute $\Delta G_{r \rightarrow rc}$, we first sample $N_L N_{\zeta}$ ligand conformations by applying samples from $U_\zeta$ to samples from the ligand MD trajectories. Then, for each sample from the protein trajectory $x_P$, we place each atom in a KD tree. We then iteratively run nearest-neighbors queries on every atom in each ligand conformation to the protein KD trees to determine the distance $d_{ij}$ between each atom $i$ in the protein and each atom $j$ in the ligand. We find all possible pairs of protein and ligand frames such that $d_{ij} \geq c_{ij}$ for all atom pairs $(i,j)$, where $c_{ij}$ is the cutoff distance between atoms $i$ and $j$. These cutoff distances are determined from the force field parameters as the interatomic distance at which the Lennard-Jones energy between $i$ and $j$ equals a cutoff energy (in this paper, 2 kcal/mol); pairs closer than $c_{ij}$ are considered steric clashes. We chose $N_\zeta = \lceil N_{\text{target}} / (N_P N_L) \rceil$, where $N_{\text{target}} = 10^9$ is the desired total number of frame pairs to test.

From these filtered samples, we determine $\Delta G_{r \rightarrow rc} = -k_B T \ln\alpha$, where $\alpha$ is the fraction of samples from $U_r$ that passed the filter. To see this, note that the partition functions of $U_r$ and $U_{rc}$ differ only in their domain of integration: $Z_{rc} = \int_{\mathcal{F}} e^{-\beta U_r(x)}\,dx$ where $\mathcal{F}$ is the clash-free region, while $Z_r = \int e^{-\beta U_r(x)}\,dx$. Thus $\Delta G_{r \rightarrow rc} = -k_BT \ln(Z_{rc}/Z_r)$, and $Z_{rc}/Z_r$ is exactly the probability that a sample drawn from $U_r$ lies in $\mathcal{F}$, which we estimate as $\alpha$.

Finally, we compute the last leg of the thermodynamic cycle $\Delta G_{rc \rightarrow 1}$ by running MBAR \cite{shirts_statistically_2008} on the samples from the protein-ligand complex trajectory, as well as the filtered samples from $U_{rc}$. If the number of samples is greater than a cutoff value $N_s$ (here, 5,000), we randomly sample $N_s$ samples. Finally, we return $\Delta G_{\text{binding}} = \Delta G_{0 \rightarrow r} + \Delta G_{r \rightarrow rc} + \Delta G_{rc \rightarrow 1}$

\section{Benchmarking DBFE}

We evaluate DBFE on two benchmarks: a host-guest benchmark of cyclodextrin and octa-acid systems, and a protein-ligand benchmark of drug-like complexes. Validating free energy methods on host-guest benchmarks is a standard practice due to their simplicity, while protein-ligand benchmarks test the ability of the algorithm to scale to real-world tasks. Both the benchmarks we used for this study were also used by \citet{setiadi_tuning_2024} in their work on optimizing force-field parameters for implicit solvent free energy calculations. For both benchmarks, we compare DBFE against Onufriev-Bashford-Case model II (OBC2) \cite{onufriev_exploring_2004} double decoupling (OBC2 DD), TIP3P \cite{jorgensen_comparison_1983} DD, and OBC2 MM/GBSA (computed from the same end-state simulations as DBFE, but without conformational entropy corrections).

To perform the OBC2 DD calculations, we used Yank \cite{noauthor_choderalabyank_2024}, which utilizes Hamiltonian replica exchange simulations \cite{wang_identifying_2013} and uses MBAR to compute the final $\Delta G$ \cite{shirts_statistically_2008}. We used ParmED \cite{noauthor_parmedparmed_2024} to remove explicit waters and add the OBC2 solvation model \cite{onufriev_exploring_2004}. Each Yank lambda window was run for 2 ns at 300K with a 2 fs Langevin integrator.

When benchmarking DBFE, we ran separate end-state simulations using OpenMM 8 \cite{eastman_openmm_2024}: receptor-only (10 independent 10 ns runs, 100 ns total), ligand-only (10 ns), and complex (2 ns). We used \texttt{pymbar} to detect equilibration and decorrelate all simulation data (with a minimum burn-in time of 0.1 ns) \cite{chodera_simple_2016}. Since both simulations could be pre-computed in a virtual screening context, we can use these longer simulations without affecting the amortized cost of DBFE. All simulations used the Amber ff14SB \cite{maier_ff14sb_2015} protein force field and GAFF 2 \cite{wang_development_2004, he_fast_2020} for small molecules, with the OBC2 implicit solvent model \cite{onufriev_exploring_2004}, at 300K with a 2 fs Langevin integrator.

We also computed OBC2 MM/GBSA binding energies from the same protein-ligand complex frames used by DBFE. MM/GBSA estimates the binding free energy by decomposing the complex potential energy into receptor, ligand, and interaction terms using the same implicit solvent model, but does not include a conformational entropy correction.

We used percentile bootstrapping with 1,000 resamples to compute 95\% confidence intervals for all metrics.

\subsection{Host-Guest Benchmark}

We evaluated DBFE on a host-guest benchmark consisting of $\alpha$-cyclodextrin (aCD), $\beta$-cyclodextrin (bCD), octa-acid (OA), and tetra-endo-methyl octa-acid (TEMOA) host systems with various small-molecule guests, drawn from the SAMPL4 and SAMPL5 host-guest challenges \cite{muddana_sampl4_2014, yin_overview_2017}. Experimental binding affinities were obtained from isothermal titration calorimetry measurements. Explicit solvent reference values were obtained from attach-pull-release (APR) calculations: cyclodextrin values from \citet{henriksen_evaluating_2017} and OA/TEMOA values from \citet{yin_overview_2017}.

For the cyclodextrin hosts, guests can bind via both the primary and secondary faces; we ran separate simulations for each pose and combined results via Boltzmann weighting across poses. After filtering complexes where the ligand left the binding site during simulation, we retained 50 host-guest complexes.

The OBC2 DD calculations used 26 complex lambda windows and 8 solvent lambda windows (34 total) per complex. In contrast, DBFE required only 3 end-state simulations (host, guest, and host-guest complex).

\begin{table}[hbt]
\centering
\caption{Comparison of methods on the host-guest benchmark (50 complexes). 95\% confidence intervals shown in brackets.}
\label{tab:mobley_models}
% \resizebox{\textwidth}{!}{%
\begin{tabular}{llll}
\toprule
Model & RMSE (kcal/mol) & Pearson $r$ & Spearman $\rho$ \\
\midrule
DD (OBC2) & 4.1 [3.4, 4.9] & 0.48 [0.19, 0.69] & 0.43 [0.12, 0.70] \\
DD (TIP3P) & 1.1 [0.87, 1.3] & 0.89 [0.77, 0.95] & 0.86 [0.75, 0.92] \\
OBC2 MM/GBSA & 13 [12, 14] & 0.31 [0.026, 0.51] & 0.25 [-0.060, 0.54] \\
\midrule
\textbf{DBFE} & 4.1 [3.4, 4.9] & 0.58 [0.33, 0.79] & 0.56 [0.28, 0.79] \\
\bottomrule
\end{tabular}
% }
\end{table}

\begin{figure}[h]
\includegraphics[width=6in]{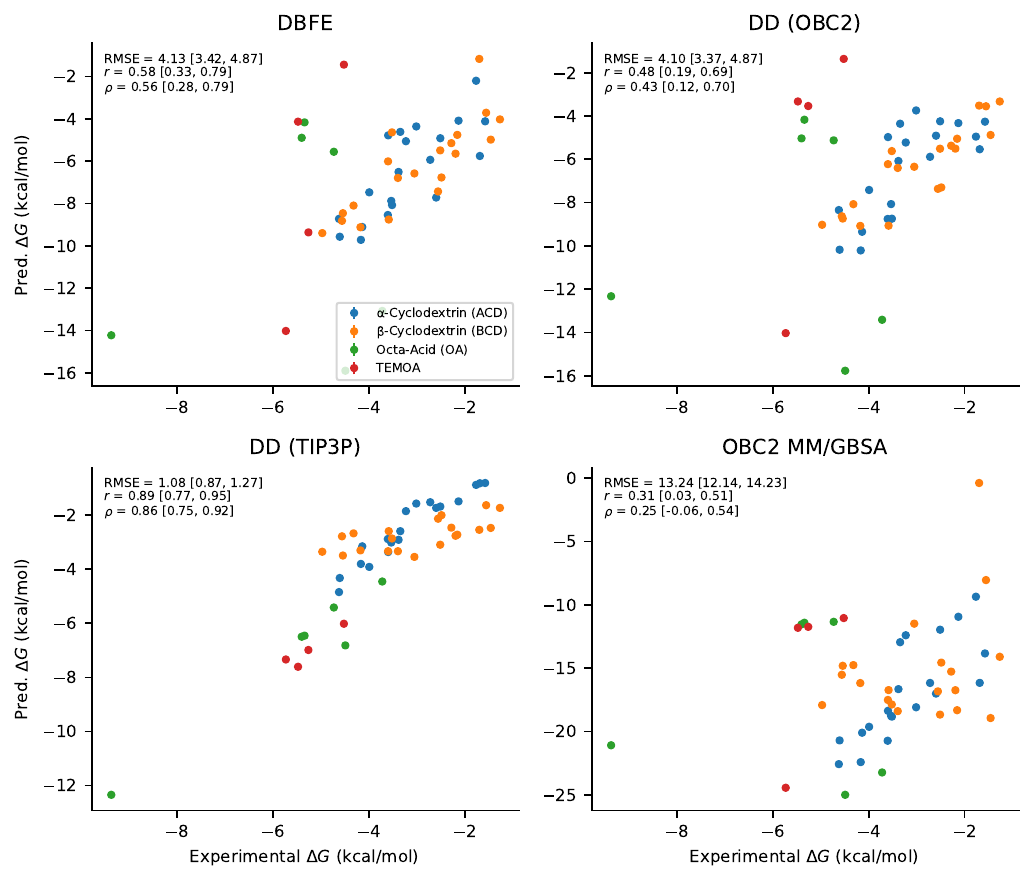}
  \caption{Comparison of all methods versus experimental $\Delta G$ on the host-guest benchmark. Each subplot shows predicted versus experimental binding free energies, with bootstrap metrics (RMSE, Pearson $r$, Spearman $\rho$) and 95\% confidence intervals.}
  \label{fig:mobley_models}
\end{figure}

The results are shown in Table \ref{tab:mobley_models} and Figure \ref{fig:mobley_models}. DBFE achieves a Pearson correlation of $r$=0.58 with experiment, outperforming OBC2 DD ($r$=0.48) on these systems. OBC2 MM/GBSA performs poorly ($r$=0.33), suggesting that the conformational entropy correction provided by DBFE is important for these host-guest systems. As expected, TIP3P DD performs substantially better ($r$=0.89), highlighting the importance of explicit water treatment even for these relatively simple systems.

\subsection{Protein-Ligand Benchmark}

We next evaluated DBFE on the protein-ligand binding affinity benchmark from \citet{alibay_evaluating_2022}. It contains experimental binding affinity annotations for 59 complexes spanning 4 protein targets (Cyclophilin D, MCL-1, HSP90, and PWWP1) with a wide range of binding affinities.

Due to steric clashes in the initial simulation geometry, we removed three complexes from the analysis (CyclophilinD:27, MCL-1:15, and PWWP1:15). We additionally excluded two complexes (MCL-1:60 and MCL-1:65) where the protein binding site closed during the protein-only simulation, preventing DBFE from finding any samples that passed the Lennard-Jones cutoff. We include previous results from \citet{alibay_evaluating_2022} for TIP3P DD calculations.

The OBC2 DD calculations used 26 complex lambda windows and 4 solvent lambda windows (30 total) per complex.

\begin{table}[hbt]
\centering
\caption{Comparison of methods on the protein-ligand benchmark (54 complexes). 95\% confidence intervals shown in brackets.}
\label{tab:fragment_models}
% \resizebox{\textwidth}{!}{%
\begin{tabular}{llll}
\toprule
Model & RMSE (kcal/mol) & Pearson $r$ & Spearman $\rho$ \\
\midrule
DD (OBC2) & 6.6 [5.8, 7.3] & 0.73 [0.63, 0.82] & 0.75 [0.60, 0.84] \\
DD (TIP3P) & 2.8 [2.4, 3.2] & 0.88 [0.81, 0.93] & 0.85 [0.75, 0.91] \\
OBC2 MM/GBSA & 28 [25, 30] & 0.71 [0.60, 0.82] & 0.70 [0.53, 0.82] \\
\midrule
\textbf{DBFE} & 5.3 [4.7, 6.0] & 0.65 [0.48, 0.80] & 0.64 [0.41, 0.81] \\
\bottomrule
\end{tabular}
% }
\end{table}

\begin{figure}[h]
\includegraphics[width=6in]{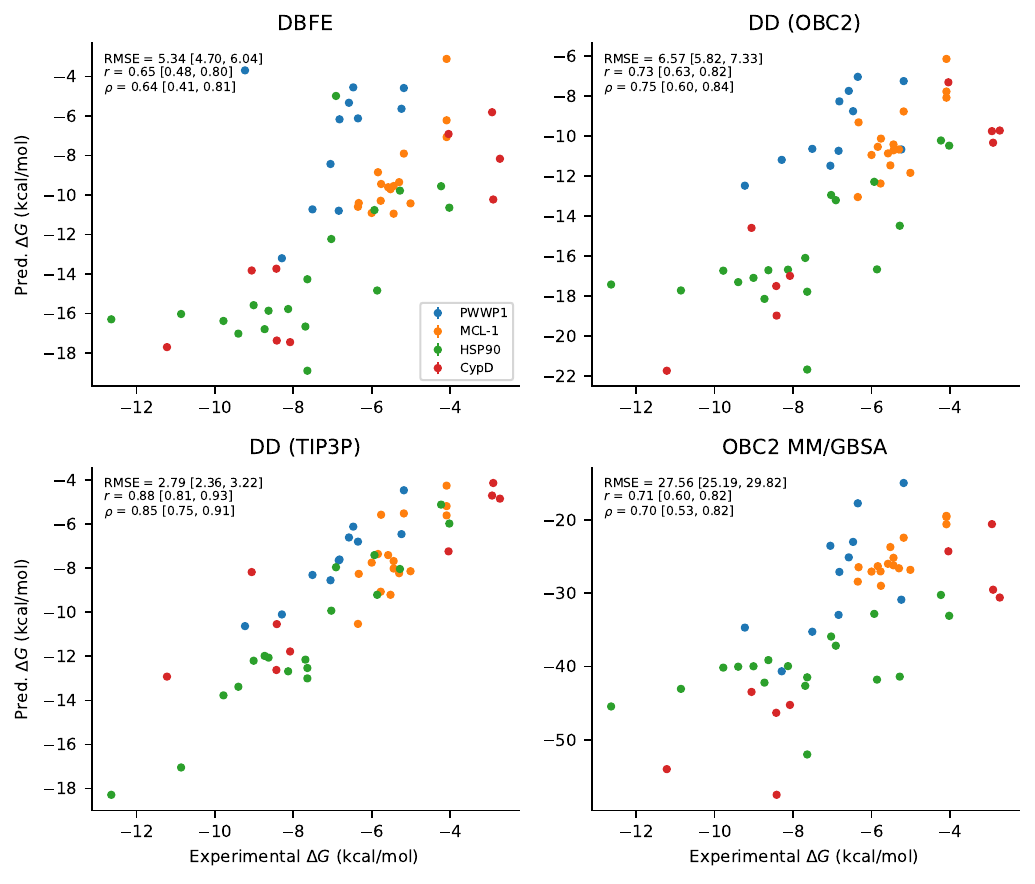}
  \caption{Comparison of all methods versus experimental $\Delta G$ on the protein-ligand benchmark. Each subplot shows predicted versus experimental binding free energies, with bootstrap metrics (RMSE, Pearson $r$, Spearman $\rho$) and 95\% confidence intervals.}
  \label{fig:fragment_models}
\end{figure}

We succeeded in running DBFE on 54 of the 56 complexes. The results are shown in Table \ref{tab:fragment_models} and Figure \ref{fig:fragment_models}. On this benchmark, DBFE achieved $r$=0.65 and $\rho$=0.64, which is slightly worse than OBC2 MM/GBSA ($r$=0.71, $\rho$=0.70). OBC2 MM/GBSA computes the binding free energy from the same implicit solvent simulations without the conformational entropy correction that DBFE provides; that it performs slightly better suggests that the conformational entropy estimate from DBFE introduces noise on these protein-ligand systems. Both methods performed worse than OBC2 DD ($r$=0.73), which in turn performed considerably worse than TIP3P DD ($r$=0.88).

\begin{figure}[h]
\includegraphics[width=6in]{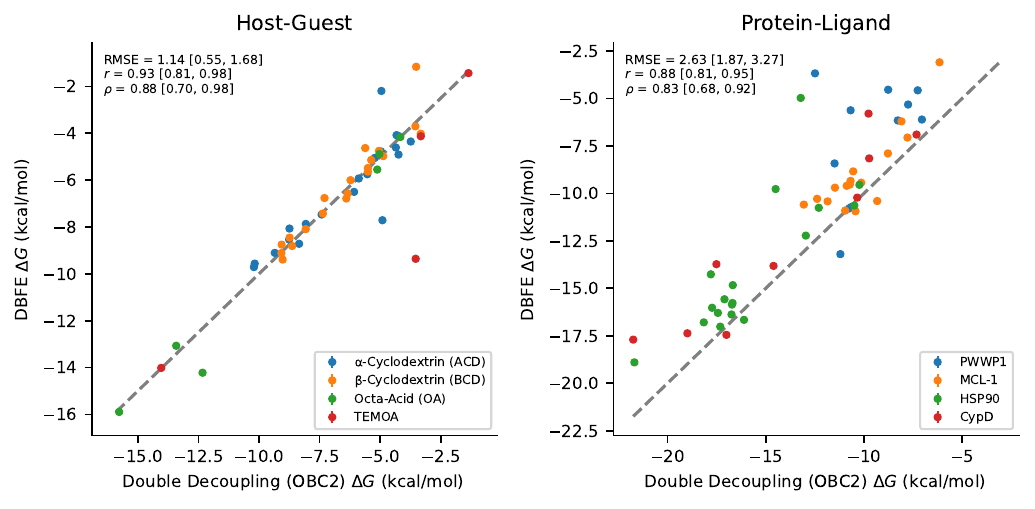}
  \caption{DBFE versus OBC2 double decoupling $\Delta G$ on both benchmarks. Left: host-guest benchmark. Right: protein-ligand benchmark.}
  \label{fig:dbfe_vs_dd}
\end{figure}

\section{Discussion}

We have introduced DBFE, a new implicit solvent ABFE method that requires only end-state simulations and no alchemical intermediates. Our results on two benchmarks reveal distinct performance characteristics depending on system complexity.

On the host-guest benchmark, DBFE ($r$=0.58) outperformed OBC2 DD ($r$=0.48) in correlation with experiment, while OBC2 MM/GBSA performed substantially worse ($r$=0.31). This suggests that for host-guest systems, the conformational entropy correction provided by DBFE is important and that the combinatorial sampling strategy of DBFE is effective for these systems with smaller, well-defined binding sites.

On the protein-ligand benchmark, DBFE ($r$=0.65) performed slightly worse than OBC2 MM/GBSA ($r$=0.71), which computes the binding free energy from the same implicit solvent simulations without conformational entropy corrections. This contrasts with the host-guest results and indicates that the conformational entropy estimate from DBFE introduces noise on these more complex protein-ligand systems. Both methods performed worse than OBC2 DD ($r$=0.73), while TIP3P DD ($r$=0.88) substantially outperformed all OBC2 methods. The gap between OBC2 and TIP3P DD ($r$=0.73 vs $r$=0.88) is much larger than the gap between DBFE and OBC2 DD ($r$=0.65 vs $r$=0.73), suggesting that improving implicit solvent models would yield greater accuracy gains than improving the free energy estimator.

DBFE's primary advantage is computational efficiency. While DBFE's total simulation time for a single complex can exceed that of DD (due to the long receptor simulation), this cost is amortized in a virtual screening context: the receptor simulation is run once and reused for all candidate ligands. The per-ligand cost of DBFE is then dominated by a single short complex simulation, compared to the 26 complex lambda windows used by OBC2 DD in our benchmark: a 26x reduction in per-ligand simulation cost. The exact speedup will vary with the DD protocol, as the number of lambda windows required depends on the system and the desired precision.

Nonetheless, DBFE has limitations. The low phase-space overlap between $U_{rc}$ and $U_1$ sometimes yields inaccurate results or causes complete failure (as with the MCL-1 complexes where the binding site closed). This could be addressed with longer receptor simulations or enhanced sampling techniques to bias toward holo-like conformations \cite{basciu_holo-like_2019}. In general, we suggest that DBFE will perform better than MM/GBSA on systems where conformational entropy differs substantially in the systems being studied and where we can sample holo-like conformations from an apo receptor simulation. On the other hand, MM/GBSA will be preferable for induced-fit systems or systems where conformational entropy is less important.

Another fundamental limitation is DBFE's reliance on implicit solvent. Advances in ML implicit solvent models \cite{katzberger_general_2024, airas_transferable_2023, chen_machine_2021} may prove vital to increasing DBFE's accuracy, particularly for protein-ligand systems where explicit water treatment is critical.

Finally, DBFE's end-state framework could be extended to \textit{relative} binding free energy (RBFE) calculations. Such a method may suffer less from phase-space overlap issues because the overlap between two ligand-bound states is typically greater than between coupled and decoupled states. This RBFE approach would not require the two ligands to share a common scaffold \cite{baumann_broadening_2023}, making it applicable to diverse chemical libraries.

\begin{acknowledgement}
The authors thank Dr. David Koes, James Wellnitz, and Travis Maxfield for insightful comments and discussion. The authors also acknowledge support from the NIH Biophysics Training Grant (T32GM148376-01A1).
\end{acknowledgement}

% \begin{suppinfo}

% This will usually read something like: ``Experimental procedures and
% characterization data for all new compounds. The class will
% automatically add a sentence pointing to the information on-line:

% \end{suppinfo}

\bibliography{references}

\end{document}